\documentclass[%
 aip,
 amsmath,amssymb,
 reprint,%
]{revtex4-1}

\usepackage{graphicx}
\usepackage{dcolumn}
\usepackage{bm}

\usepackage[utf8]{inputenc}
\usepackage[T1]{fontenc}
\usepackage{mathptmx}
\usepackage[dvipsnames]{xcolor}

\begin{document}

\preprint{AIP/123-QED}

\title[Optical and dielectric properties of MoO$_3$ nanosheets for van der Waals heterostructures]{Optical and dielectric properties of MoO$_3$ nanosheets for van der Waals heterostructures}

\author{Daniel Andres-Penares}%
\affiliation{Institute of Photonics and Quantum Sciences, SUPA, Heriot-Watt University, Edinburgh EH14 4AS, U.K.
}%
\affiliation{ICMUV, Instituto de Ciencia de Materiales, Universidad de Valencia, P.O. Box 22085, 46071 Valencia, Spain
}%
\author{Mauro Brotons-Gisbert}%
\author{Cristian Bonato}%
\affiliation{Institute of Photonics and Quantum Sciences, SUPA, Heriot-Watt University, Edinburgh EH14 4AS, U.K.
}%
\author{Juan F. Sánchez-Royo}%
\affiliation{ICMUV, Instituto de Ciencia de Materiales, Universidad de Valencia, P.O. Box 22085, 46071 Valencia, Spain
}%
\affiliation{MATINÉE: CSIC Associated Unit-(ICMM-ICMUV of the University of Valencia), Universidad de Valencia, P.O. Box 22085, 46071 Valencia, Spain
}%
\author{Brian D. Gerardot}%
 \email{Author to whom correspondence should be addressed: B.D.Gerardot@hw.ac.uk}%
\affiliation{Institute of Photonics and Quantum Sciences, SUPA, Heriot-Watt University, Edinburgh EH14 4AS, U.K.
}%

\date{\today}

\begin{abstract}
Two-dimensional (2D) insulators are a key element in the design and fabrication of van der Waals heterostructures. They are vital as transparent dielectric spacers whose thickness can influence both the photonic, electronic, and optoelectronic properties of 2D devices. Simultaneously, they provide protection of the active layers in the heterostructure. For these critical roles, hexagonal Boron Nitride (hBN) is the dominant choice due to its large bandgap, atomic flatness, low defect density, and encapsulation properties. However, the broad catalogue of 2D insulators offers exciting opportunities to replace hBN in certain applications that require transparent thin layers with additional optical degrees of freedom. Here we investigate the potential of single-crystalline Molybdenum Oxide (MoO$_3$) as an alternative 2D insulator for the design of nanodevices that require precise adjustment of the light polarization at the nanometer scale. First, we measure the wavelength-dependent refractive indices of MoO$_3$ along its three main crystal axes and determine the in-plane and out-of-plane anisotropy of its optical properties. We find the birefringence in MoO$_3$ nanosheets compares favorably with other 2D materials that exhibit strong birefringence, such as black phosphorus, ReS$_2$, or ReSe$_2$, in particular in the visible spectral range where MoO$_3$ has the unique advantage of transparency. Finally, we demonstrate the suitability of MoO$_3$ for dielectric encapsulation by reporting linewidth narrowing and reduced inhomogeneous broadening of 2D excitons and optically active quantum emitters, respectively, in a prototypical monolayer transition-metal dichalcogenide semiconductor. These results show the potential of MoO$_3$ as a 2D dielectric layer for manipulation of the light polarization in vertical 2D heterostructures.
\end{abstract}

\maketitle

The remarkable diversity of van der Waals (vdW) layered crystals, the breadth of their properties, and the ability to stack them without restriction has positioned atomically-thin crystals as a unique platform to develop novel heterostructure-based devices with engineered functionalities \cite{geim2013van}. 
A key ingredient in the design of vdW heterostructures for electronic and photonic applications is an insulator: a transparent layered dielectric material to encapsulate the active material for both protection and device functionality. Among the broad catalog of wide-bandgap materials offered by the vdW family, high-quality hexagonal Boron Nitride (hBN) is the dominant choice due to its atomic flatness and low defect density. Encapsulation of graphene or transition metal dichalcogenides (TMDs) with hBN can significantly reduce environmental disorder \cite{Rhodes2019disorder} to yield significantly enhanced transport \cite{dean2010boron,bandurin2017high} and optical \cite{cadiz2017excitonic,ajayi2017approaching, Wierzbowski2017} properties. In addition, hBN is an invaluable dielectric spacer for engineering the optical properties  \cite{Gerber2018, brotons2018engineering, Fang2019, Raja2017, Unuchek2019} and can act as a tunnel barrier between active 2D regions \cite{brotons2019coulomb, Shimazaki2020atacMLhBN, Pan2018} with a tunnelling probability that can be precisely controlled by the number of atomic layers \cite{britnell2012electron}. 
While hBN is ubiquitous due to these exceptional properties, some drawbacks exist. 
High-purity hBN is typically obtained only in extreme conditions \cite{Watanabe2004}, creating scarcity. Further, the nuclear spins of all isotopes of both boron and nitrogen are non-zero \cite{P1992, Stone2014}. Suppl. Table S1 presents the nuclear spin number (\textit{I}) and natural abundance for each isotope of boron and nitrogen. A zero nuclear spin number can only be obtained when both atomic numbers \textit{Z} and \textit{A} are even for all atoms in a material, a requirement that hBN fails to meet. For some applications in quantum information science, this leads to spin decoherence, as observed in magnetic resonance spectroscopy of atomically thin hBN \cite{Lovchinsky2017}. This can negatively affect the spin coherence of an electron localized at a lattice defect \cite{Gottscholl2020} or a quantum dot encapsulated by hBN \cite{brotons2019coulomb, Baek2020, Klein2021}. 

Beyond a passive role as an isotropic dielectric material, an insulating vdW material that can protect the active region of a heterostructure \textit{and} play additional roles in the device functionality provides new opportunities. For instance, due to the extreme sensitivity of excitons in 2D materials such as monolayer (ML) transition metal dichalcogenides (TMDs) to their environment, it is possible to control and tune the electronic and optical properties (such as the bandgap and exciton binding energies) using dielectric engineering techniques  \cite{Raja2017}. Further, specific hBN thicknesses can be chosen for particular applications to optimise specific characteristics, e.g., to engineer the far-field radiation pattern \cite{brotons2018engineering} or to control the tunnel barrier for carriers and interlayer coupling \cite{Gerber2018, brotons2019coulomb, Shimazaki2020atacMLhBN}. But, in the case of hBN, its refractive index is isotropic within the exfoliation plane making thickness the only variable to control in the heterostructure design. In this context, in-plane anisotropic insulators could offer a new knob to tune the dielectric properties. 
This degree of freedom could prove valuable for next-generation optoelectronic, photonic, and quantum devices.

In this framework, an intriguing alternative to hBN is single-crystalline MoO$_3$. MoO$_3$ has a relatively large indirect bandgap (> 3 eV \cite{hussain2003characterization,balendhran2013two}), ideal for applications requiring a transparent material in the visible and near-infrared part of the spectrum. For example, MoO$_3$ has been used in silicon \cite{Geissbuhler2015}, perovskite \cite{Zhao2014}, and organic \cite{Tseng2012} photovoltaics applications and as a buffer layer in organic light-emitting diodes \cite{shin2008bulk}. Moreover, the dominant isotopes of both molybdenum and oxygen have zero nuclear spins (see Suppl. Table S1), positioning MoO$_3$ as a promising candidate for applications that require spin coherence. In addition, MoO$_3$ presents an in-plane anisotropic crystal structure in its layered phase ($\alpha$-MoO$_3$) \cite{zhang2017high,Molina-Mendoza2016,Wei2020}, enabling a platform for ultra-low-loss polaritons due to anisotropy \cite{Ma2018} and anisotropic phonon-polariton propagation along the surface of a rotated MoO$_3$ bilayer \cite{Hu2020}. Similar crystal anisotropies have been exploited to fabricate optical and optoelectronic devices \cite{yuan2015polarization,wang2017short,yang2017optical}, positioning MoO$_3$ as a candidate for nanodevices in which transparent thin layers with an additional tunable variable are needed. Although the dielectric properties of single-crystal bulk $\alpha$-MoO$_3$ and its biaxial optical anisotropy have been measured by valence electron-energy-loss spectroscopy \cite{Lajaunie2013} and interferometric methods \cite{deb1968physical}, the wavelength-dependent in-plane and out-of-plane birefringence of single-crystalline MoO$_3$ nanoflakes and the performance MoO$_3$ as an encapsulation dielectric for vdW heterostructures has not yet been investigated.

Here, we characterize the optical and encapsulating properties of MoO$_3$ nanosheets. First, using an imaging ellipsometer, we measure the refractive indices along the \textit{a}, \textit{b}, and \textit{c} axes and determine the in-plane and out-of-plane wavelength-dependent birefringence of the refractive indices for MoO$_3$. To unambiguously demonstrate the in-plane anisotropy, we measure the polarization dependent reflectivity of a MoO$_3$ nanosheet and observe the effective refractive index is dependent on the incident polarization. Building on these results, we calculate the optical phase retardance as a function of sample thickness and wavelength of these MoO$_3$ nanosheets. Finally, we apply MoO$_3$ as an encapsulating material for a ML WSe$_2$ layer. We probe the absorption at cryogenic temperatures of the WSe$_2$ A-exciton and, relative to a reference unencapsulated ML WSe$_2$ sample, observe substantial linewidth narrowing and an energetic blue-shift. We ascribe this improvement to the atomic flatness, reduced disorder, and dielectric environment of the MoO$_3$. Further, we characterize a single quantum emitter in the MoO$_3$ encapsulated WSe$_2$, and observe linewidths on par with those observed in hBN encapsulated WSe$_2$. The performance of MoO$_3$ as an encapsulation dielectric is found to be comparable to hBN. Overall, we find the intrinsic anisotropy of the optical properties of MoO$_3$ provides an unexplored degree of freedom useful for photonic applications while simultaneously MoO$_3$'s dielectric properties suggest its suitability as a candidate to replace hBN in certain applications. 

We prepare MoO$_3$ nanosheets via mechanical exfoliation from bulk crystals (2DSemiconductors, Scottsdale, AZ, USA) in an Ar environment. The thickness of the exfoliated nanosheets is estimated \textit{in situ} by optical contrast analysis (see Suppl. Figure S1), a reliable and accurate method to characterize the thickness of 2D materials \cite{Brotons-Gisbert2017, Krecmarova2019}, including MoO$_3$ \cite{Puebla2020}. The MoO$_3$ nanosheets employed as encapsulating layers in our vertical vdW heterostructures are 15 - 40 nm thick, while the MoO$_3$ layers studied via ellipsometry and differential reflectivity measurements are 110 - 150 nm thick.

Assembly of heterostructures incorporating WSe$_2$ (2DSemiconductors, Scottsdale, AZ, USA) were carried out using the hot pick-up transfer technique \cite{Pizzocchero2016} in the inert environment to promote pristine interfaces. Spectroscopic ellipsometry measurements were carried out using a spectroscopic imaging (spatial resolution $\sim$1 $\mu$m) nulling ellipsometer (EP4, Accurion Gmbh, Göttingen, Germany) in an Ar atmosphere at room temperature in the wavelength range from 410 nm to 980 nm with constant wavelength steps of 10 nm. Optical reflectance and photoluminescence (PL) spectroscopy measurements were carried out at T=4K using a diffraction limited confocal microscope.

\begin{figure}
\includegraphics[width=1\columnwidth]{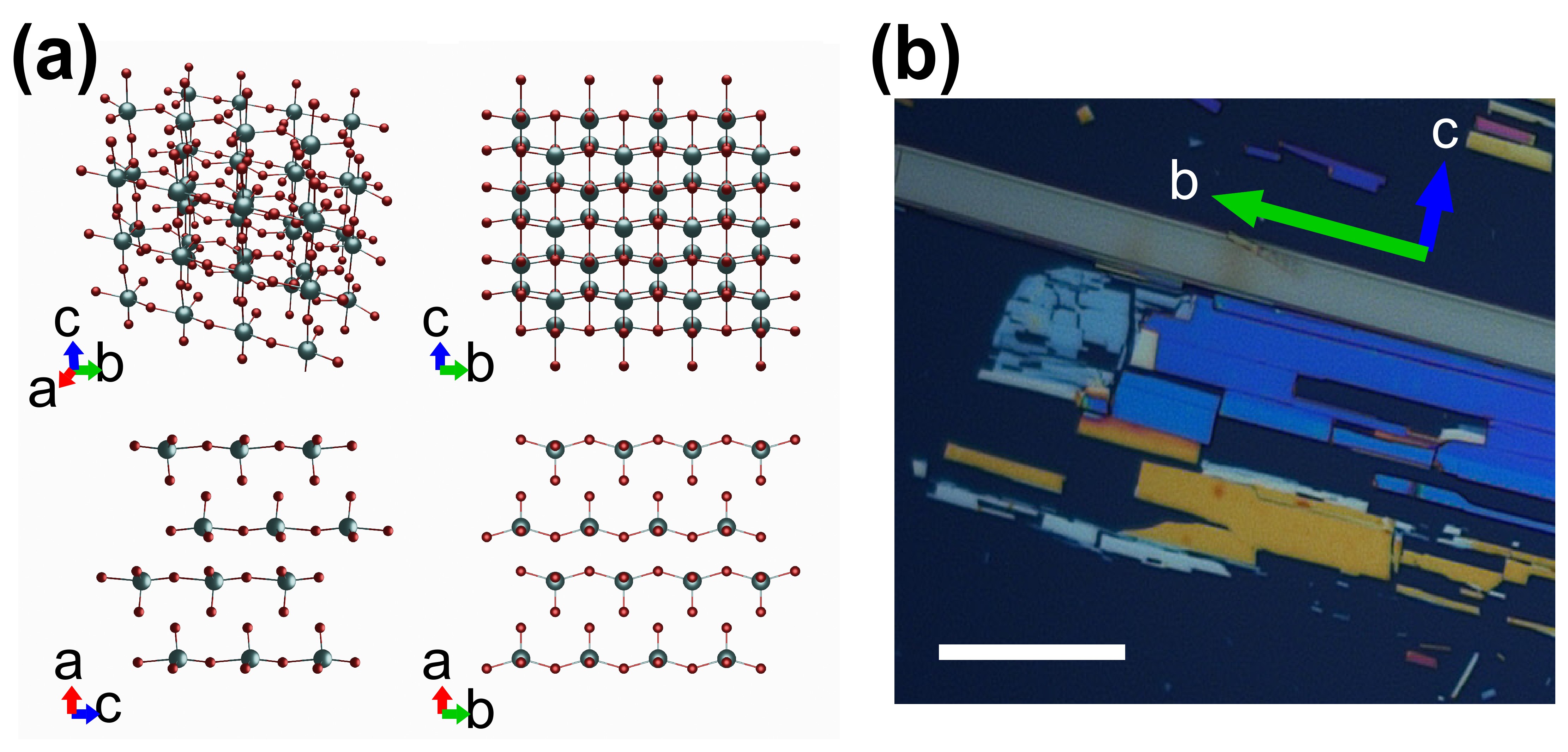}
\caption{\label{fig:F1} (a) Orthorhombic $\alpha$ - MoO$_3$ (space group Pnma62) structure showing the standard orientation of the crystal shape and views along the $a$, $b$ and $c$ axis. (b) Optical images of exfoliated MoO$_3$ nanosheets. Scale bar = 20 $\mu$m.}
\end{figure}

$\alpha$ - MoO$_3$, with a centrosymmetric Pnma (62) structure \cite{Kim2019, Yao2012} as shown in Figure \ref{fig:F1}(a), is the thermally stable phase of MoO$_3$. It has a laminar structure with lattice parameters $a = 3.761$ \r{A}, $b = 3.969$ \r{A} and $c = 14.425$ \r{A} ($\alpha = \beta = \gamma = 90^{\circ}$) \cite{Lajaunie2013,Kim2019}. Mo - O bonding within each layer is covalent whereas bonding between layers is of vdW type, making possible its exfoliation (Figure \ref{fig:F1}(b)). However, within the in-plane crystallographic directions, a drastic difference in bond strength can be observed both in Figure \ref{fig:F1}(a), comparing views from $b$ and $c$ axis, and in Figure \ref{fig:F1}(b), which reveals clear rectangular shapes. The directions defined by different bonding strengths make it possible to identify the direction parallel to longer sides (higher bond density) and shorter ones (weaker bond density) with $b$ and $c$ directions, respectively, while $a$ is the out-of-plane direction. 

\begin{figure}[t]
\includegraphics[width=0.8\columnwidth]{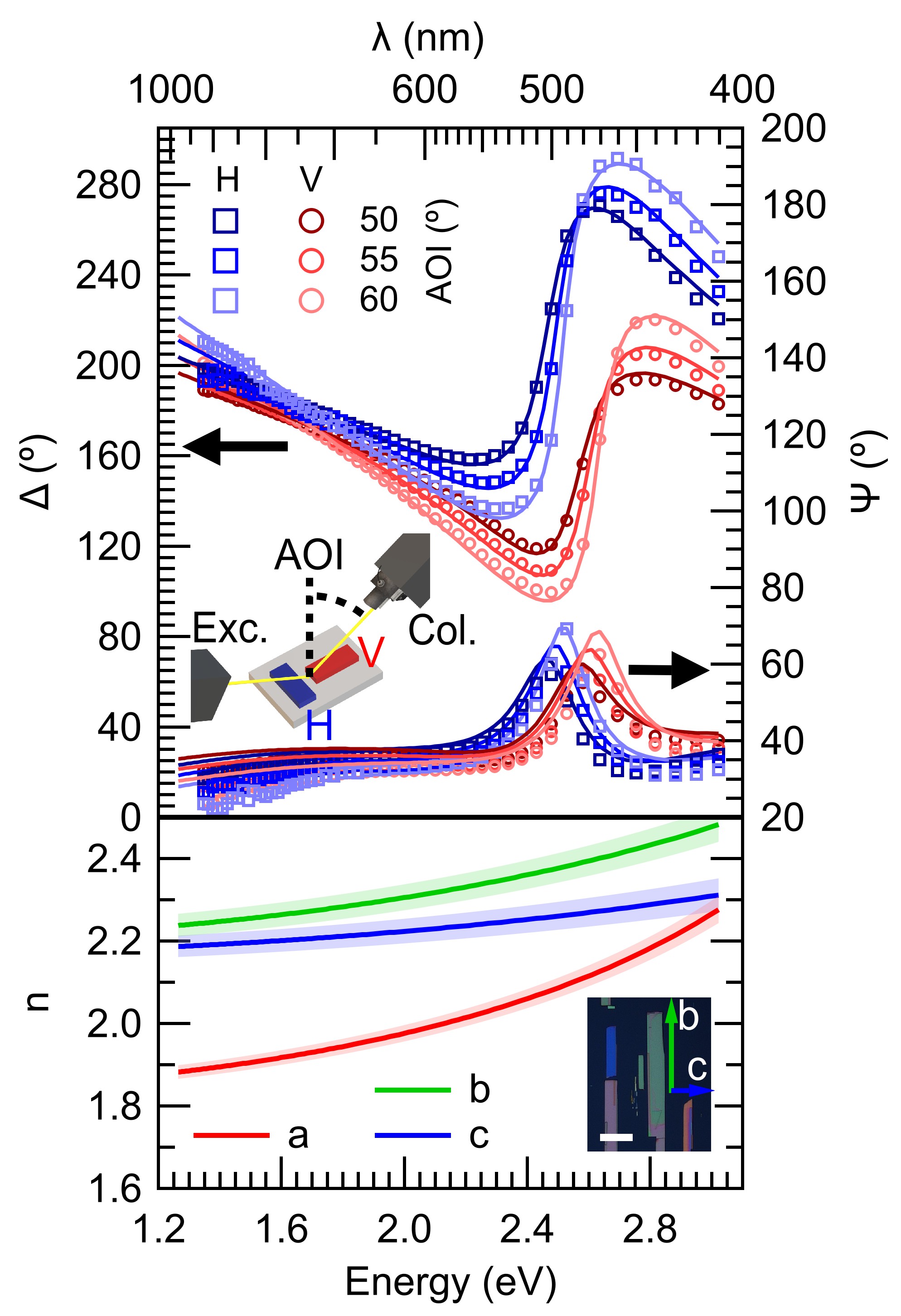}
\caption{\label{fig:F2} $\Delta$ and $\Psi$ measured for three different incident angles and two different sample orientations (as described in the inset in the top panel) via ellipsometry of an MoO$_3$ nanosheet (top panel) and the refractive indices along the \textit{a}, \textit{b} and \textit{c} axis obtained from fits using a Sellmeier dispersion law (bottom panel). The inset in the bottom panel identifies the crystal axes. Scale bar = 20 $\mu$m.}
\end{figure}

To probe and directly correlate the optical properties with crystal anisotropy, we perform imaging spectroscopic ellipsometry of MoO$_3$ samples. Figure \ref{fig:F2}, top panel, shows the ellipsometric angles $\Delta$ and $\Psi$ (open circles and squares) measured for a 112$\pm1$ nm thick MoO$_3$ nanosheet on a Si/SiO$_2$ substrate. Ellipsometric data from the samples were acquired at three different angles of incidence (defining the vertical as $AOI = 0^{\circ}$) in five-degree steps and two orthogonal incidence planes obtained via two sets of measurements taken by rotating the samples in the azimuthal angle $\theta$ = 90$^\circ$ within the $b$-$c$ plane (in-plane). In this scenario, the wave vector of the excitation beam is contained within the $a$-$b$ crystallographic plane (labeled as H configuration, in blue tones, when the longer side of the nanosheets --$b$ crystallographic direction-- is perpendicular to the direction described by the incidence and nanosheet planes) or within the $a$-$c$ plane (labeled as V configuration, in red tones, when the longer side of the nanosheets --$b$ crystallographic direction-- is parallel to the direction described by the incidence and nanosheet planes). For a quantitative analysis of the optical properties of the MoO$_3$ nanosheet it is necessary to fit the measured data with an appropriate multilayer model. With the aim of minimizing substrate-induced uncertainties in the determination of the refractive index of MoO$_3$ flakes, ellipsometric data from the bare Si/SiO$_2$ substrate were also measured simultaneously on a spot very close to the MoO$_3$ nanosheet using the same experimental conditions. Suppl. Figure S2 shows the ellipsometric angles $\Delta$ and $\Psi$ measured for the bare Si/SiO$_2$ substrate, from which an oxide thickness of 97.3 $\pm$ 0.1 nm was determined by employing the reported refractive indices of Si and SiO$_2$ \cite{Henrie2004}. Due to the strong crystal anisotropy of MoO$_3$, we modelled its refractive index as a biaxial birefringent medium, where $n_a$, $n_b$  and $n_c$ are the refractive indices along each crystal axis. Because the extinction coefficients for MoO$_3$ in the visible range are very low \cite{Puebla2020}, we neglect that term in our model and assume real-valued refractive indices described by a Sellmeier dispersion law of the form $n_i^2(\lambda) = 1+B_i\lambda^2/(\lambda^2-C_i)$, with $i = a, b, c$. The solid lines in the top panel in Figure \ref{fig:F2} represent numerical fits of the experimental data to our theoretical model, from which the refractive indices along the different crystallographic axes are obtained (see the bottom panel in Figure \ref{fig:F2}). The good agreement between the experimental values and the theoretical fits verifies that our assumption of real-valued refractive indices for MoO$_3$ nanoflakes in the visible range is reasonable. The Sellmeier coefficients resulting from the numerical fits are summarized in Suppl. Table S2. As can be seen in the bottom panel in Figure \ref{fig:F2}, the refractive index of MoO$_3$ nanosheets presents a clear biaxial anisotropy, with values of $n_a$, $n_b$ and $n_c$ in good agreement with previously reported values for single-crystalline bulk MoO$_3$ \cite{deb1968physical} and a recent study on MoO$_3$ nanosheets \cite{Puebla2021}. The results in Figure \ref{fig:F2} also highlight a large in-plane/out-of-plane birefringence in MoO$_3$ for visible wavelengths. At $\lambda$ = 633 nm we measure $\Delta n_{ac}=n_a-n_c\sim-0.25$ and $\Delta n_{ab}=n_a-n_b\sim-0.33$, which are higher in absolute value than the values found in other birefringent crystals such as YVO$_4$ ($\Delta n = 0.21$ at 633 nm \cite{shi2001measurement}) which is widely used in laser devices and optical components.

In addition to the in-plane/out-of-plane birefringence, MoO$_3$ nanosheets also show a marked in-plane birefringence, for instance $\Delta n_{bc}=n_b-n_c\sim 0.11$ at 520 nm. Although smaller than the in-plane birefringence of black phosphorus ($\Delta n=0.245$ at 520 nm \cite{yang2017optical}, birefringence that has been recently explored for its potential use in reconfigurable color displays \cite{Jia2021}), the measured in-plane birefringence for MoO$_3$ is $\sim$3 times larger than that of other vdW materials such as ReS$_2$ ($\sim$0.037 \cite{yang2017optical}) and ReSe$_2$ ($\sim$0.047\cite{yang2017optical}) at 520 nm, and is comparable to that of state-of-the-art bulk birefringent materials such as CaCO$_3$ ($\sim$0.17\cite{yang2017optical}). However, unlike the previously mentioned vdW materials with large in-plane birefringence, MoO$_3$ is transparent in the visible spectral range \cite{Puebla2021}, opening this feature to applications where a transparent material is necessary. Such in-plane birefringence positions MoO$_3$ as an ideal candidate for polarization-integrated vdW nanodevices that require precise adjustment of the light polarization at the nanometer scale. The electric field components of a plane wave propagating along the $a$ axis of MoO$_3$ with a linear polarization state misaligned with respect to the $b$ and $c$ crystallographic axes experience a phase retardance $\delta$ given by $\delta (d,\lambda) = 2\pi\Delta n_{bc}(\lambda)d/\lambda$, with $d$ being the MoO$_3$ thickness. Since the band structure of MoO$_3$ has a negligible thickness dependence \cite{Molina-Mendoza2016}, we do not expect the refractive index to depend on the nanosheet thickness. Therefore, for a given $\lambda$ the phase retardance $\delta$ grows linearly with the sample thickness. Given that the ML thickness of MoO$_3$ is 0.7 nm \cite{cai2017rapid}, we estimate a maximum phase retardance of $\sim0.105^{\circ}$ per atomic layer at 410 nm. Suppl. Figure S3 shows the calculated phase retardance induced by MoO$_3$ as a function of the sample thickness and light wavelength.

\begin{figure}[t]
\includegraphics[width=0.9\columnwidth]{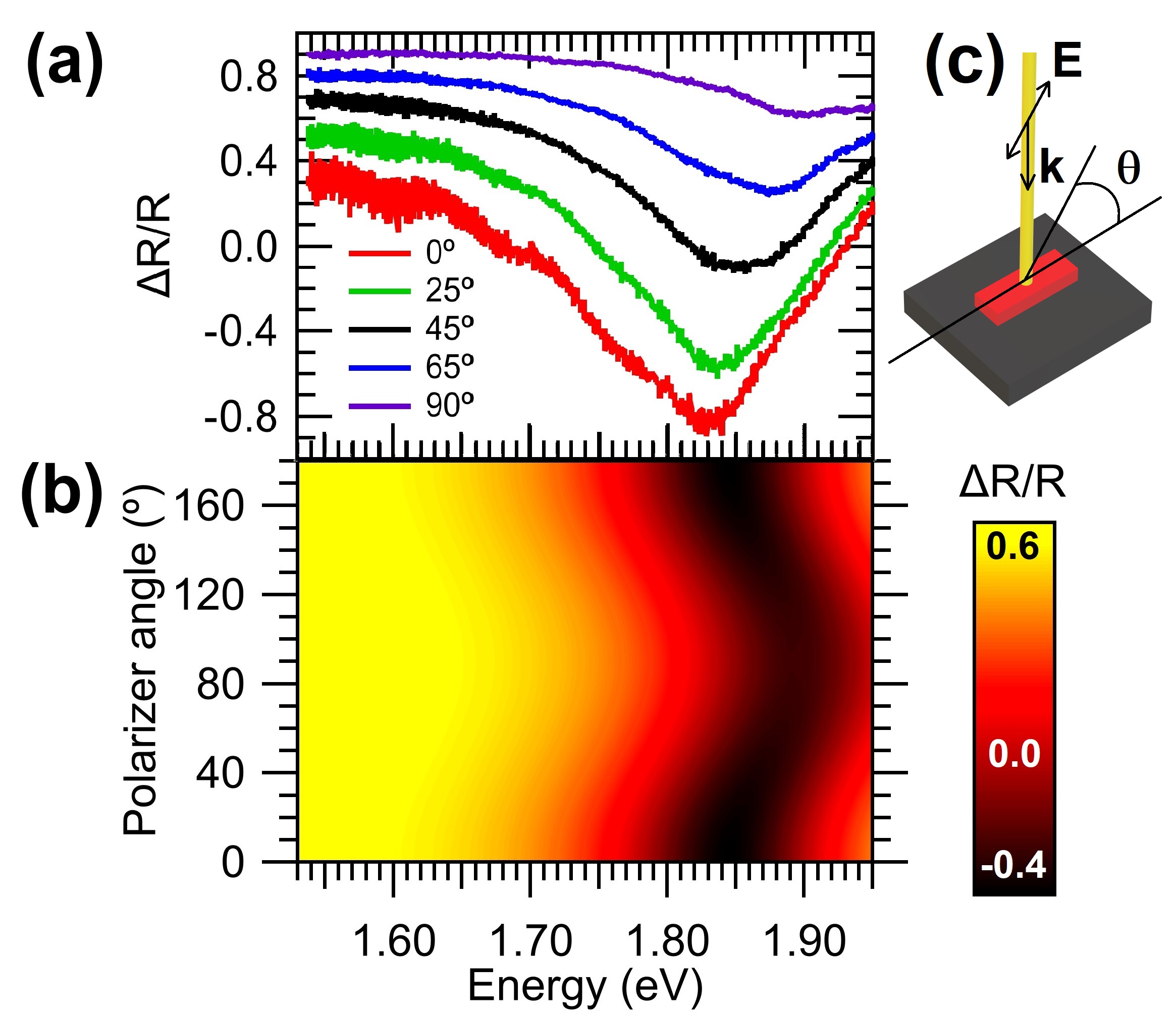}
\caption{\label{fig:F3} (a) Experimental $\Delta R/R$ spectra and (b) calculated using a transfer matrix method rotating in-plane a MoO$_3$ nanosheet, from a parallel position with a collection polarizer (0º, vertical nanosheet, $b$ axis) to a perpendicular one (90º, horizontal nanosheet, $c$ axis). (c) Schematic representation of the azimuthal angle ($\theta$) between the nanosheet orientation and the collection polarizer.}
\end{figure}

To further show the role that the in-plane anisotropy of MoO$_3$ can play in the design of nanodevices for photonics applications, we perform linear-polarization-resolved optical reflectance spectroscopy measurements in MoO$_3$ samples exfoliated on top of a Si/SiO$_2$ substrate. During the measurements, the azimuthal angle of the linearly polarized incident light is rotated with respect to the crystallographic in-plane axes of MoO$_3$ with steps of 5 degrees in a vertical diffraction limited confocal microscope with normal incidence ($AOI = 0^{\circ}$). Figure \ref{fig:F3}(a) shows linear reflectance contrast spectra ($\Delta R/R$) acquired for a MoO$_3$ nanosheet under different incident angles, where $\theta = 0^{\circ}$ and $\theta = 90^{\circ}$ represent incident linearly-polarized light aligned with the $b$ and $c$ axis, respectively. The linear reflectance contrast spectra were calculated as $\Delta R/R = (R_s - R_0)/R_s$, with $R_s$ and $R_0$ the reflectivity measured in the sample and the bare substrate, respectively. The $\Delta R/R$ spectra show clear minima that blueshift when the incident light is rotated from parallel to the $b$ axis to parallel to the $c$, in addition to an overall intensity change. The origin of such minima can be attributed to the destructive interference enabled by the optical thickness of MoO$_3$, for which the reflected intensity $R_s$ becomes minimum at a wavelength $\lambda=4dn_{eff}(\theta)/(2m+1)$, with $d$ being the MoO$_3$ thickness, $n_{eff}(\theta)$ being an effective in-plane refractive index, and $m$ being the order of the interference maximum. Due to the in-plane anisotropy of MoO$_3$, the incident polarized light experiences a different refractive index $n_{eff}(\theta)$ which depends on the azimuthal angle $\theta$ through the equation:
\begin{align}
    \frac{1}{n^2_{eff}(\theta)}=\frac{\cos^2{\theta}}{n^2_b}+\frac{\sin^2{\theta}}{n^2_c},
    \label{neff}
\end{align}
which states that the optical thickness for which the constructive interference is satisfied depends on the azimuthal angle of the incident light. This result yields a redshift of the wavelength of the $\Delta R/R$ minimum as the incident angle is rotated from $0^{\circ}$ to $90^{\circ}$. Figure \ref{fig:F3}(b) shows the calculated $\Delta R/R$ of a $153\pm2$ nm MoO$_3$ nanosheet on top of a Si/SiO$_2$ substrate with an oxide layer of 97 nm as a function of the incident angle. The calculations were carried out using a multilayer transfer matrix using the in-plane refractive indices shown in Figure \ref{fig:F2}. These results not only reproduce the observed blueshift of the $\Delta R/R$ minima but also account for the observed intensity change of the $\Delta R/R$ as function of the illumination angle.

\begin{figure}
\includegraphics[width=1\columnwidth]{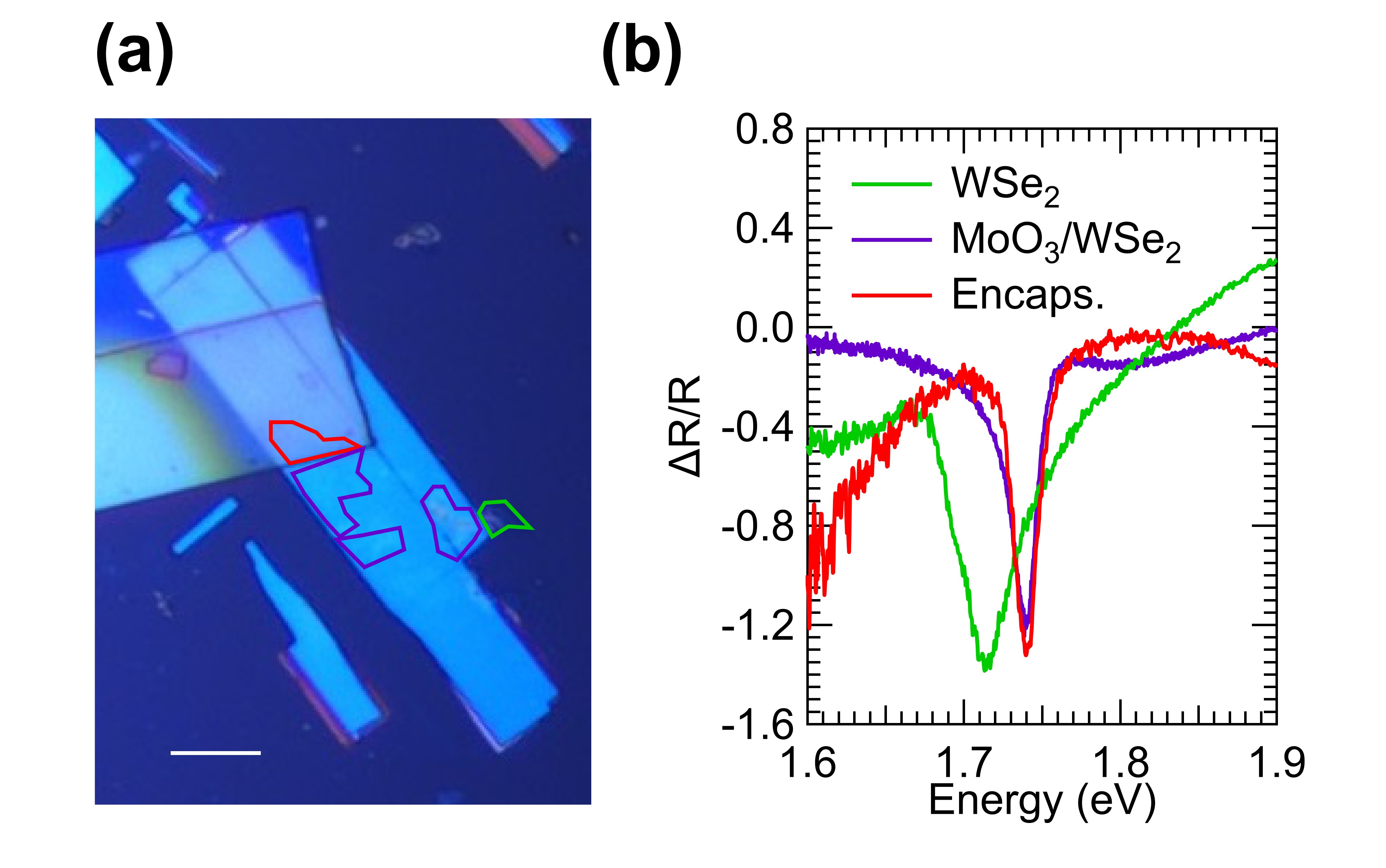}
\caption{\label{fig:F4} (a) Optical image of a ML WSe$_2$ (outlined with solid lines) with three distinct dielectric environments: sitting directly on a Si/SiO$_2$ substrate (green area), sitting directly on top of MoO$_3$ (purple area), and fully encapsulated by MoO$_3$ (red area). (b) $\Delta R/R$ comparing the ML WSe$_2$ from the three regions described in (a). Scale bar in (a) represents 10 $\mu$m.}
\end{figure}

Next, to investigate the potential of MoO$_3$ as a dielectric spacer and encapsulating layer for vdW heterostructure devices, we fabricated a sample consisting of a ML WSe$_2$ partially encapsulated by MoO$_3$ on top of a Si/SiO$_2$ (90 nm) substrate. Figure \ref{fig:F4}(a) shows an optical image of the fabricated heterostructure, in which three different regions can be identified: i) a region where the WSe$_2$ ML lays directly on top of the Si/SiO$_2$ substrate (area outlined in green); ii) a region where the WSe$_2$ ML is separated from the Si/SiO$_2$ substrate by a bottom MoO$_3$ layer (purple area); and iii) a region where the WSe$_2$ ML is fully encapsulated by top and bottom MoO$_3$ layers (red area). Figure \ref{fig:F4}(b) shows low-temperature (LT) $\Delta R/R$ spectra measured at the spatial positions indicated in Figure \ref{fig:F4}(a) in the energy range 1.6-1.9 eV. In this spectral region, the fundamental direct optical transition (so-called A-excitons at the $K$ and $K'$ points of the hexagonal Brillouin zone) dominates the optical response of the WSe$_2$ ML. We observe the WSe$_2$ exciton resonances in the partially (purple) and fully encapsulated (red) regions show a pronounced shift of $\sim$26 meV to higher energies as compared to the WSe$_2$ on SiO$_2$. Energy shifts of similar magnitude have also been observed for WSe$_2$ and other 2D TMDs deposited on top of different substrates and encapsulated by hBN or layers with different refractive indices \cite{Lin2014,stier2016probing,Raja2017,Raja2019,Steinleitner2018}. The observed energy shifts are the result of the local dielectric screening of the Coulomb interactions induced by the dielectric surroundings of the 2D layers \cite{Lin2014, Raja2017,Raja2019,Steinleitner2018}, which leads to substantial and opposite shifts of the free-carrier bandgap $E_g$ and the exciton binding energy $E_B$. Although for encapsulated samples both magnitudes might change considerably, in our experiments we only can access the exciton energy (i.e. the difference between $E_g$ and $E_B$). Therefore, the resulting energy shift $\Delta E$ (either to higher or lower energies), is the result of the relative energy shifts $\Delta E_g$ and $\Delta E_B$, which typically give rise to a very small shift compared to the exciton energy in non-encapsulated samples. In our case, the observed energy blueshift for the exciton resonance indicates that $\Delta E_g > \Delta E_B$. Moreover, the observed energy shift of the exciton resonance is accompanied by a substantial narrowing of the exciton linewidth for the partially and fully encapsulated regions of the sample ($\sim13$ meV) as compared to the non-encapsulated one ($\sim40$ meV). Similar narrowing of exciton linewidths, attributed to a reduction of the inhomogeneous broadening, has also been observed for different ML TMDs encapsulated by hBN as a consequence of protection from possible substrate-related charge and electric field fluctuations as well as increased flatness enabled by the hBN encapsulation \cite{cadiz2017excitonic,ajayi2017approaching}. Although linewidths approaching the homogeneous limit (as narrow as $\sim$2 meV) have been reported, the exciton linewidths we measure for WSe$_2$ encapsulated by MoO$_3$ are very similar to the ones we routinely obtain for hBN-encapsulated WSe$_2$ (see Suppl. Figure S4). Further improvements could be possible with higher purity WSe$_2$ crystals \cite{Edelberg2019} and incorporation into electrostatically gated devices. These results confirm that MoO$_3$ represents an alternative to hBN as a dielectric spacer and protecting layer for vdW-based heterostructure devices. 

\begin{figure}
\includegraphics[width=1\columnwidth]{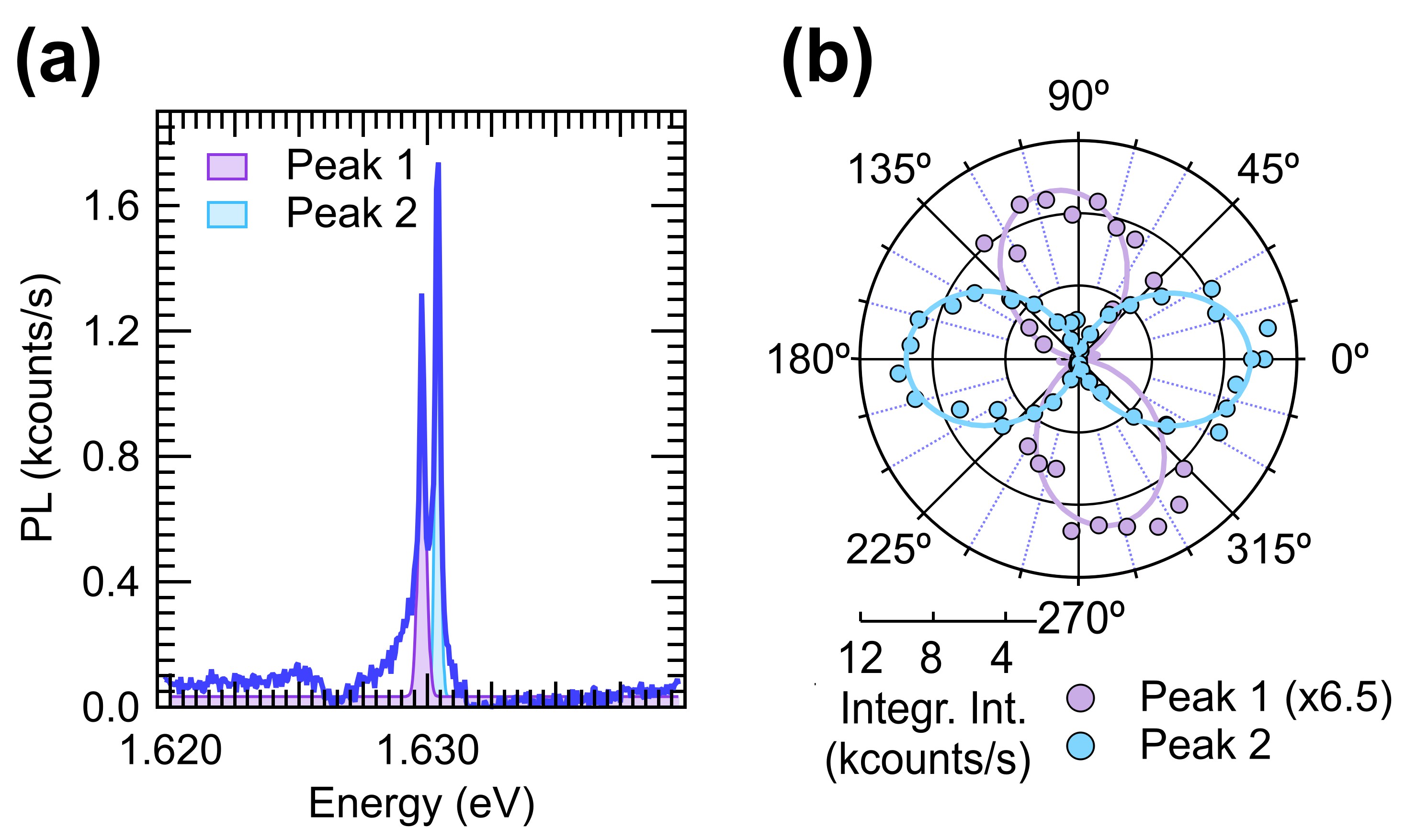}
\caption{\label{fig:F5} (a) PL of a single quantum emitter in a MoO$_3$ encapsulated WSe$_2$ sample at T = 4K. The spectrum shows a fine structure splitting of $600$ $\mu$eV with peak linewidths of $\sim$ 150 and $\sim$ 230 $\mu$eV. (b) Polarization dependent spectroscopy of the doublet the peaks have orthogonal linear polarization.}
\end{figure}

Finally, motivated by potential applications for MoO$_3$ in quantum devices, we investigate the effect MoO$_3$ encapsulation has on the properties of quantum emitters in ML WSe$_2$. Confocal PL imaging of several bare MoO$_3$ flakes at a temperature of 4K confirms the absence of LT emission from MoO$_3$ in the energy range 1.55-1.95 eV. Conversely, confocal PL scanning of MoO$_3$-encapsulated ML WSe$_2$ reveals a few localized spots with higher PL intensity than the homogeneous emission background. These localized bright spots present discrete spectrally narrow peaks originating from WSe$_2$ quantum emitters \cite{brotons2019coulomb,srivastava2015optically,he2015single,tonndorf2015single, Kumar2015}. Figure \ref{fig:F5}(a) shows a PL spectrum of the neutral exciton of a single WSe$_2$ quantum emitter fully encapsulated in MoO$_3$. The spectrum exhibits a clear doublet, split by $600$ $\mu$eV, with orthogonally linear polarized emission, as shown in Figure \ref{fig:F5}(b). The emission doublet is a fine-structure splitting arising from the electron–hole exchange interaction energy and the asymmetry in the confinement potential \cite{srivastava2015optically,he2015single,tonndorf2015single,brotons2019coulomb, Kumar2015}. The fine-structure doublet exhibits the typical saturation behavior of a single quantum emitters with increasing power, as shown in Suppl. Figure S5. The individual fine-structure split peaks exhibit emission linewidths of $\sim$ 150 and $\sim$ 230 $\mu$eV and minimal spectral fluctuation at long timescales. Over 10 minutes, the standard deviation of the peak energy is $\sim$20 $\mu$eV (see Suppl. Figure S6). Overall, the inhomogeneous broadening of the WSe$_2$ quantum emitters fully encapsulated in MoO$_3$ is comparable to those fully encapsulated in hBN \cite{brotons2019coulomb}.

In summary, we have demonstrated that MoO$_3$ is a viable alternative to hBN as a 2D dielectric insulator, in particular for quantum devices requiring spin coherence or applications requiring an extra degree of freedom to fine-tune photonic heterostructures due its in-plane birefringence. We report the refractive indices along the three crystal axes of MoO$_3$ and find that the in-plane birefringence of MoO$_3$ is among the highest reported for a 2D material. These results are complemented by differential reflectivity measurements. As an example advantage of its unusually high birefringence, MoO$_3$'s potential use as an optical phase retardance element is proposed. Finally, we demonstrated the suitability of MoO$_3$ nanosheets for dielectric encapsulation, reporting linewdith narrowing in ML WSe$_2$ and reduced inhomogeneous broadening of optically active TMD quantum emitters. These results pave the way for MoO$_3$ as a dielectric element for vertical vdW heterostructures.

\section*{Supplementary Material}
The following are available online. Table S1: Nuclear spin number (\textit{I}) and abundance in nature of hBN and MoO$_3$ constituents: Boron, Nitrogen, Oxygen and Molybdenum, Figure S1: Optical contrast analysis for MoO$_3$ nanosheets thickness identification, Figure S2: Ellipsometry $\Delta$ and $\Psi$ (circled dots) measured on the bare Si/SiO$_2$ substrate, Table S2: Sellmeier coefficients resulting from the numerical fits in a dispersion law of the form $n_i^2(\lambda) = 1+B_i\lambda^2/(\lambda^2-C_i)$, with $i = a, b, c$ being the axes in MoO$_3$ nanoflakes described in Figure 2, Figure S3: Calculated phase retardance induced by a MoO$_3$ nanosheet as a function of the sample thickness and light wavelength, Figure S4: Typical $\Delta R/R$ in WSe$_2$ monolayer encapsulated in hBN at LT, showing a narrowing similar to the obtained in MoO$_3$ encapsulated WSe$_2$ monolayer in Figure 4, Figure S5: Fine structure splitting power dependence in LT emitters in an encapsulated WSe$_2$ ML on MoO$_3$, Figure S6: Jittering in the emitter in a WSe$_2$ ML encapsulated in MoO$_3$ described in Figure 5

\begin{acknowledgments}

This   work   is   supported   by   the   EPSRC   (grant   no. EP/P029892/1),  the  ERC  (grant  no. 725920) and the EU Horizon 2020 research and innovation program under grant agreement no. 820423. B.D.G. is  supported  by  a  Wolfson  Merit  Award  from  the  Royal Society and a Chair in Emerging Technology from the Royal Academy of Engineering. M.B.-G. is supported by a Royal Society University Research Fellowship. D.A.-P. acknowledges fellowship no. UV-INV-PREDOC17F1-539274 under the program "Atracció de Talent, VLCCAMPUS" of the University of Valencia for its funding through a research stay.

\section*{Data Availability Statement}
The data that support the findings of this study are available from the corresponding author upon reasonable request.

\end{acknowledgments}

\bibliography{bib_MoO3}

\end{document}